\documentclass{article}
\usepackage{spconf,amsmath,graphicx,hyperref}
\usepackage{booktabs,makecell}
\usepackage[dvipsnames]{xcolor}
\usepackage{bm}
\usepackage{subcaption}

\title{Stream-Voice-Anon: Enhancing Utility of Real-Time Speaker Anonymization via Neural Audio Codec and Language Models}
\name{Nikita Kuzmin$^{1,2*}$\thanks{${}^{*}$Equal contribution.}, Songting Liu$^{1*}$, Kong Aik Lee$^3$, Eng Siong Chng$^1$ }
\address{
  $^1$Nanyang Technological University, Singapore \\
  $^2$Institute for Infocomm Research, A$^\star$STAR, Singapore \\
  $^3$The Hong Kong Polytechnic University, Hong Kong \\
 s220028@e.ntu.edu.sg, lius0114@e.ntu.edu.sg
 }

\begin{document}
\ninept
\maketitle
\setlength{\abovedisplayskip}{3pt} 
\setlength{\belowdisplayskip}{3pt} 
\setlength{\belowcaptionskip}{-5pt}
\addtolength{\parskip}{-0.5mm}

\begin{abstract}
Protecting speaker identity is crucial for online voice applications, yet streaming speaker anonymization (SA) remains underexplored. Recent research has demonstrated that neural audio codec (NAC) provides superior speaker feature disentanglement and linguistic fidelity. NAC can also be used with causal language models (LM) to enhance linguistic fidelity and prompt control for streaming tasks. However, existing NAC-based online LM systems are designed for voice conversion (VC) rather than anonymization, lacking the techniques required for privacy protection. Building on these advances, we present \emph{Stream-Voice-Anon}, which adapts modern causal LM-based NAC architectures specifically for streaming SA by integrating anonymization techniques. Our anonymization approach incorporates pseudo-speaker representation sampling, a speaker embedding mixing and diverse prompt selection strategies for LM conditioning that leverage the disentanglement properties of quantized content codes to prevent speaker information leakage. Additionally, we compare dynamic and fixed delay configurations to explore latency-privacy trade-offs in real-time scenarios. Under the VoicePrivacy 2024 Challenge protocol, \emph{Stream-Voice-Anon} achieves substantial improvements in intelligibility (up to 46\% relative WER reduction) and emotion preservation (up to 28\% UAR relative) compared to the previous state-of-the-art streaming method \emph{DarkStream} while maintaining comparable latency (180ms vs 200ms) and privacy protection against lazy-informed attackers, though showing 15\% relative degradation against semi-informed attackers.

\end{abstract}
\section{Introduction}
\label{sec:intro}

Speaker anonymization (SA) transforms speech to conceal speaker identity while preserving usability for downstream tasks such as automatic speech recognition (ASR) or speech emotion recognition (SER). Streaming anonymization presents additional challenges due to strict latency constraints that require real-time processing with minimal delay. Furthermore, the source speaker's identity is typically unknown beforehand, preventing offline anonymization strategies that rely on speaker-specific adaptations, thus necessitating speaker-agnostic approaches.

Offline SA approaches can be categorized into digital signal processing (DSP) methods \cite{9306379, patino21_interspeech, Tavi_2022, mawalim22_spsc} and deep learning (DL) \cite{10096607, 10244064, kuzmin24_spsc} approaches. DSP methods are computationally cheap, but provide limited privacy and controllability. DL approaches are divided into conventional pipelines using ASR to TTS cascades \cite{10022601} or voice conversion (VC) \cite{champion2023anonymizing} methods based mostly on continuous features, and neural audio codec (NAC)-based \cite{10447871, panariello24_spsc, yao24_spsc} approaches leveraging discrete acoustic tokens. Conventional DL methods suffer from speaker information re-entanglement during vocoding. In contrast, NAC-based methods encode speech into quantized representations using learned codebooks, enabling language model (LM) architectures to generate speech while maintaining clearer separation of linguistic content from speaker characteristics.

Real-time anonymization is crucial for call centers, voice assistants, live legal recordings, and sensitive medical conversations \cite{meyer2025use}. Existing streaming approaches \cite{10832303, quamer2025darkstream} mostly adapt conventional offline designs rather than leveraging modern LM architectures, inheriting speaker information re-entanglement issues and requiring privacy-enhancing techniques like k-means clustering that degrade utility. In contrast, LM-based streaming VC has progressed rapidly \cite{wang2024streamvoice}, with causal LMs enabling stable, low-latency zero-shot conversion while inheriting superior disentanglement traits of offline NAC methods.

Inspired by advances in streaming VC, we adopt \emph{Stream-Voice-Anon}, a real-time voice anonymization system that uses LM-based streaming VC architectures. A causal streaming content encoder with VQ bottleneck extracts speaker-invariant content tokens by distilling self-supervised learning features from HuBERT \cite{2021ITASL..29.3451H, zhang2024speechtokenizer}. A causal autoregressive LM converts these tokens into synthesized acoustic codes by conditioning on pseudo-speaker embeddings and content with acoustic tokens extracted from various prompt utterances. The model employs a dual decoder architecture \cite{fish-speech-v1.4}: a primary autoregressive transformer generates frame-level latent representations, while a secondary lightweight transformer reconstructs multiple codebook layers. The system incorporates configurable frame delays (1-8 tokens) to balance latency and quality, achieving latencies as low as 180\,ms on a 3060 RTX laptop GPU. Evaluation using the VoicePrivacy 2024 Challenge framework \cite{tomashenko2024voiceprivacy} demonstrates that \emph{Stream-Voice-Anon}\footnote{Our demo page: \href{https://paniquex.github.io/Stream-Voice-Anon}{https://paniquex.github.io/Stream-Voice-Anon} } provides privacy protection comparable to the SOTA streaming pipeline \emph{DarkStream} \cite{quamer2025darkstream} while significantly enhancing speech quality.

\section{Proposed Approach}
\begin{figure*}[t]
  \centering
  \begin{subfigure}{0.49\textwidth}
    \includegraphics[width=\linewidth]{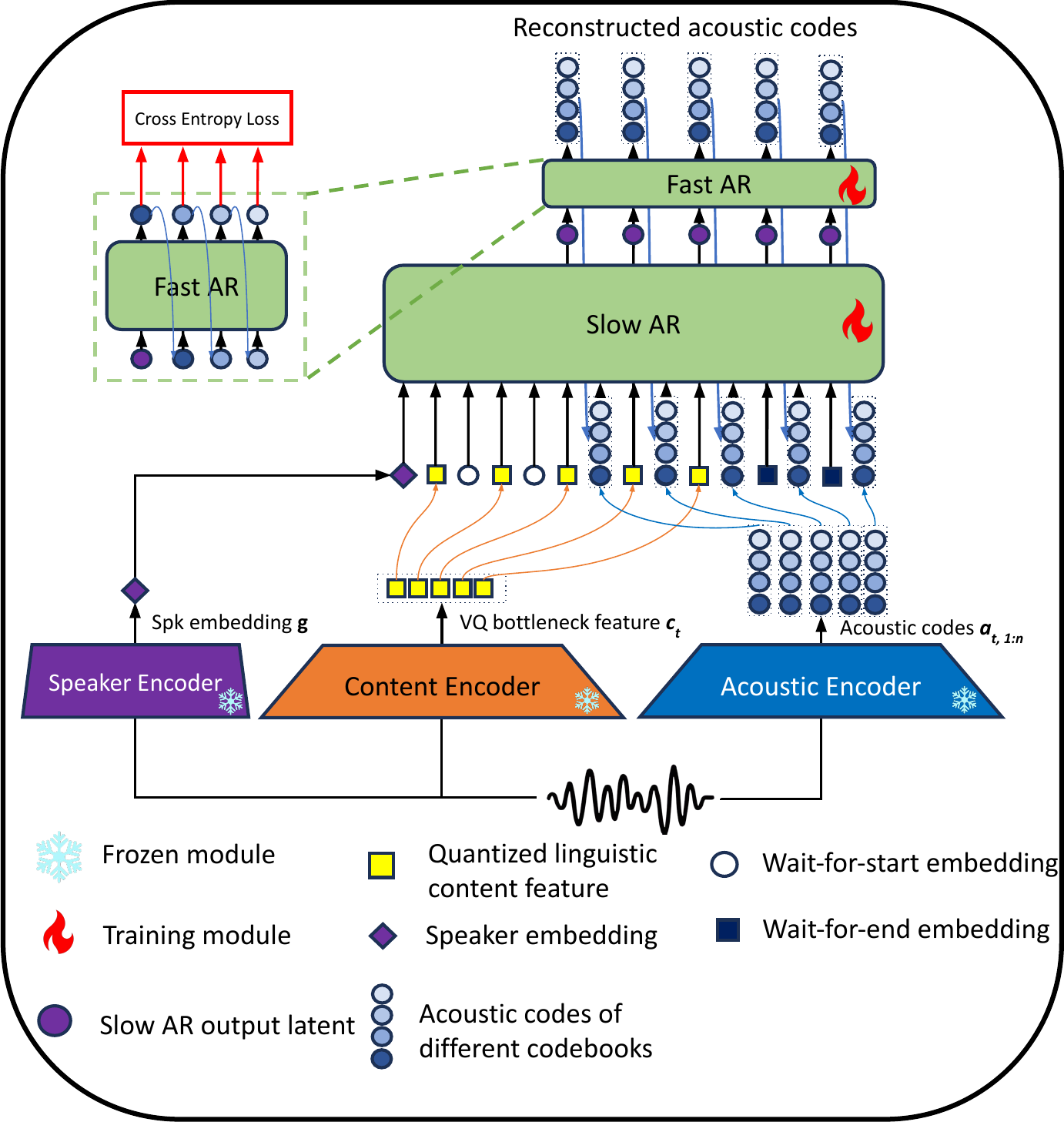}
    \caption{Training}
  \end{subfigure}\hfill
  \begin{subfigure}{0.49\textwidth}
    \includegraphics[width=\linewidth]{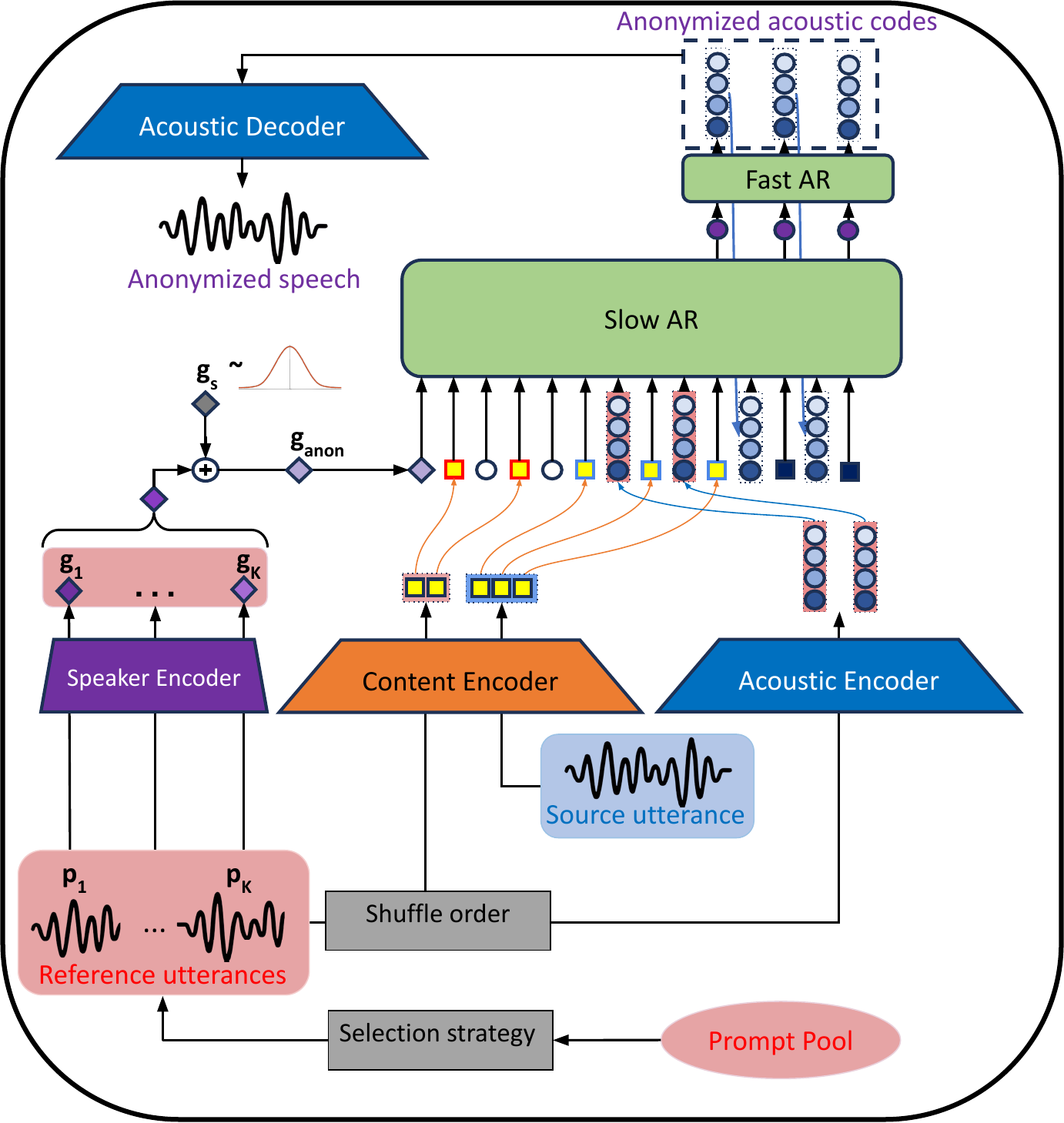}
    \caption{Inference}
  \end{subfigure}
  \caption{Training and inference pipelines of the \textbf{\emph{Stream-Voice-Anon}}.}
  \label{fig:streamvoiceanon}
\end{figure*}

\vspace{-4pt}\subsection{Overview}
The schematic of the proposed approach is illustrated in Figure \ref{fig:streamvoiceanon}. The framework consists of a content encoder, an acoustic encoder, a speaker encoder, an autoregressive voice conversion (ARVC) model based on a dual-stage transformer architecture (Slow AR and Fast AR), and a prompt pool with selection strategy for pseudo-speaker conditioning during inference. The training and inference pipelines illustrate how these components work together to enable real-time SA while maintaining linguistic content and emotion preservation. The architecture is detailed in subsequent sections.

\vspace{-0.4cm}\subsection{Content encoder}
\label{subsec:streaming-content-encoder}
We implement the content encoder ($C_e$) which extracts speaker-invariant content tokens $\{c_t\}_{t=0}^{T-1}$ using strictly causal feature extraction with zero look-ahead, following StreamVoice \cite{wang2024streamvoice}. Unlike StreamVoice, we train an ASR encoder from scratch in parallel to guide bottleneck features and employ distillation on HuBERT \cite{zhang2024speechtokenizer} representations. A VQ layer \cite{van2017neural} on the encoder output discretizes causal states using a learned codebook to obtain quantized tokens. This VQ bottleneck reduces residual mutual information with the source speaker while preserving phonetic content. $C_e$ is frozen during the AR voice conversion training.

\vspace{-0.4cm}\subsection{Acoustic encoder}
The acoustic encoder ($A_e$) processes the audio utterance and extracts multi-codebook acoustic tokens $\{a_{t,1:n}\}_{t=0}^{T-1}$ that represent the acoustic characteristics of the speech. Each frame $t$ contains $n$ acoustic codes from different codebooks, capturing various aspects of the audio signal. $A_e$ is pre-trained and frozen during the training stage of AR voice conversion.

\vspace{-0.4cm}\subsection{Speaker encoder}
For each training sample, we extract a speaker embedding $g$ using a pre-trained speaker verification model ($S_e$) and insert it at the beginning of the Slow AR sequence as a global condition.

\vspace{-4pt}\subsection{Autoregressive Voice Conversion (ARVC)}
\textbf{Feature preparation.}
We select the $A_e$ whose frame rate matches that of $C_e$.
Given a single training utterance, we then extract aligned sequences of \emph{quantized content tokens} $\{c_t\}_{t=0}^{T-1}$ from $C_e$, \emph{acoustic tokens} $\{a_{t,1:n}\}_{t=0}^{T-1}$ from $A_e$ and a global speaker embedding $g$ from $S_e$.

\textbf{Interleaved AR factorization.}
Following StreamVoice \cite{wang2024streamvoice}, to respect streaming I/O, we model the joint distribution with an interleaved AR ordering over \emph{frames}: at the frame level we alternate content and acoustics,
\[
g,\, c_0,\, a_{0, 1:n},\, c_1,\, a_{1, 1:n},\, c_2,\, a_{2, 1:n},\, \ldots,
\]
meaning the model \emph{consumes} $c_t$ from the source stream and \emph{emits} the corresponding acoustic codes $a_{t,1:n}$ of the same frame before moving on. 
This interleaving mirrors real-time VC: the source speech is read frame-by-frame, and the converted acoustics are produced frame-synchronously.

\textbf{Latency--quality trade-off via delayed emission.}
Because $c_t$ is a heavily compressed (speaker-invariant) representation, emitting $a_t$ \emph{immediately} after seeing $c_t$ is challenging without any look-ahead.
StreamVoice+ \cite{wang2024streamvoice+} introduced a fixed frame-level \emph{delay} so that the model first accumulates the future content context and only then starts emitting acoustics. For example, with $d{=}2$ the interleaving becomes
\[
g,\, c_0,\, \langle\mathrm{w4s}\rangle,\; c_1,\, \langle\mathrm{w4s}\rangle,\; c_2,\, a_{0, 1:n},\; c_3,\, a_{1, 1:n},\; c_4,\, a_{2, 1:n},\; \ldots
\]
where $\langle\mathrm{w4s}\rangle$ is a learned \emph{wait-for-start} embedding occupying output steps until emission begins. However, this fixed-delay approach lacks flexibility during inference, as the model cannot dynamically adjust the quality-latency trade-off based on real-time requirements. To address this, we introduce the dynamic-delay technique by sampling $d\!\sim\!\mathcal{U}\{1,\ldots,8\}$ per utterance, which teaches the decoder to operate under variable look-ahead. At inference time, $d$ can be chosen flexibly to trade latency for recognition fidelity (smaller WER with larger $d$).

\textbf{Two-stage AR with per-frame codebook decoding.} As each acoustic token $a_{t,1:n}$ consists of multiple codebooks, predicting all $n$ codebooks of a frame with a single AR is suboptimal. Inspired by FishSpeech \cite{fish-speech-v1.4}
We therefore use a \emph{two-stage} decoder:
a \emph{Slow AR} operates at the frame rate to produce a latent $z_t$ after receiving $c_t$ (and previous outputs), and a lightweight \emph{Fast AR} runs \emph{within} the frame conditioned on $z_t$ to decode all $n$ codebooks autoregressively:
\[
z_t,\, a_{t,1},\, a_{t,2},\, \ldots,\, a_{t,n}.
\]
Training uses teacher forcing inside the frame and the overall loss is the sum of cross-entropy over all codebooks and frames,
\[
\mathcal{L}_{\text{AR}}=\sum_{t=0}^{T-1}\sum_{k=1}^{n}\mathrm{CE}\!\left(\hat{a}_{t,k},\, a_{t,k}\right),
\]
where $\hat{a}_{t,k}$ and $a_{t,k}$ predicted logits and ground truth labels correspondingly.

\vspace{-4pt}\subsection{Inference-time anonymization techniques}
In addition to the training-time disentanglement techniques described above that help protect privacy, we strengthen anonymization with an inference-time speaker embedding mixing scheme and prompt-based randomization as illustrated in Figure \ref{fig:streamvoiceanon}(b). First, we select $K$ utterances from the prompt pool $P$ using one of our selection strategies. Before extracting tokens for the target utterances, we shuffle the selected prompts in random order and then extract both content and acoustic tokens from each prompt, concatenating them together to form diverse acoustic contexts. We use extracted representations to condition ARVC.

In parallel, we extract speaker embeddings $\{g_\text{i}\}_{i=1}^K$ from selected prompt utterances using $S_e$ and compute their average. Inspired by \cite{yao24_spsc}, we then sample a speaker embedding $g_\text{s}$ from a Gaussian distribution and form the anonymized target embedding as a linear combination:
\[
g_\text{anon} = \alpha \frac{1}{K}\sum^K_{i=1} g_\text{i} + (1 - \alpha)g_\text{s},
\]
where $\alpha$ indicates the trade-off between staying close to the prompts and injecting randomness for even stronger privacy. 

It is worth noting, all of the anonymization steps can be performed independently of source speaker utterance, thus allowing to precompute all representations beforehand and use during streaming inference.

\section{Experimental Setup}

\vspace{-4pt}\subsection{Datasets}
For training, we use LibriHeavy \cite{kang2024libriheavy}, CommonVoice \cite{commonvoice:2020}. 
The content encoder requires paired speech--text supervision; we therefore use the official transcripts of LibriHeavy and tokenize them with the \textit{OpenAI Whisper} tokenizer to form target sequences for the ASR auxiliary. The ARVC model is trained on the same dataset as the content encoder but it does not use transcripts.
The prompt pool ($P$) comprises VCTK \cite{Yamagishi2019CSTRVC}, ESD \cite{9413391}, VoxCeleb1 \cite{Nagrani17}, CREMA-D \cite{6849440}.

For evaluation, we use LibriSpeech dev-clean and test-clean to assess privacy and ASR utility, and the IEMOCAP development and evaluation sets to assess emotion recognition utility. 

\vspace{-4pt}\subsection{Evaluation Protocol}

We follow the 2024 VoicePrivacy Challenge protocol \cite{tomashenko2024voiceprivacy}. Our first privacy setting is the lazy-informed attacker ($\text{ASV}_\text{eval}$): an automatic speaker verification model based on ECAPA-TDNN \cite{Desplanques_2020}, trained on original LibriSpeech train-clean-360. Additionally, we report results for a semi-informed attacker ($\text{ASV}_\text{eval}^\text{anon}$) using the same architecture trained LibriSpeech train-clean-360 that has been anonymized at the utterance level to match the anonymization system. Automatic speech recognition $\text{ASR}_\text{eval}$ and speech emotion recognition $\text{SER}_\text{eval}$ models are pretrained on the original LibriSpeech train-960 and IEMOCAP datasets respectively. We report EER for privacy, WER for intelligibility, and UAR for emotion preservation; higher EER and UAR are better, and lower WER is better. All scores are computed using the official VPC2024 evaluation pipeline \cite{tomashenko2024voiceprivacy}.

\vspace{-4pt}\subsection{Model Configurations}

\textbf{Content encoder.}
We use a \emph{streaming} content encoder that converts 44.1\,kHz audio to 160-bin log-Mel features and downsamples by 4 times, yielding $\sim$\,21.5\,Hz latents. 
The backbone is a lightweight ConvNeXt \cite{liu2022convnet} stack followed by an 8-layer decoder-only Transformer (model/FFN dims 512/1536, SwiGLU \cite{shazeer2020glu}, RoPE \cite{su2024roformer}). 
A VQ layer with a vocabulary of 8192 entries produces the discrete content codes. 
All convolutions are causal to enable streaming.

\textbf{Acoustic encoder and decoder.}
A Firefly-GAN vocoder from FishSpeech~\cite{fish-speech-v1.4} synthesizes audio from the acoustic codes. 
Its frame rate is aligned with the content encoder ($\sim$\,21.5\,Hz) and it uses 8 codebooks. 
Both encoder and decoder components employ causal convolutions for streaming synthesis.

\textbf{Speaker encoder.} Pre-trained CAM++ \cite{wang23ha_interspeech} and SparkTTS global tokenizer \cite{wang2025sparktts} are used to extract speaker embeddings.

\textbf{ARVC.}
We adopt a two-stage ARVC: a Slow-AR (12 layers, 768 hidden dimension/2304 feed-forward dimension) produces a per-frame state that conditions a Fast-AR (4 layers, 768 hidden dimension/2304 feed-forward dimension), which then autoregressively emits all 8 acoustic codebooks for that frame.

\textbf{Training details.}
Both the content encoder and the autoregressive voice conversion (ARVC) model were trained on 8 NVIDIA H100 GPUs for 400,000 optimization steps. We used the AdamW optimizer with an initial learning rate of \(1.0\times10^{-4}\), decayed exponentially every optimization step. The batch size was 16 per GPU (effective batch size \(= 8 \times 16 = 128\)).

\textbf{Anonymization details.} We use $\alpha=0.9$ in formula for $g_\text{anon}$. Our prompt pool $P$ consists of 4 datasets.
We employ several selection strategies to evaluate both anonymization effectiveness and emotion preservation capabilities:\emph{vctk-1fix} selects one utterance from a fixed VCTK speaker; \emph{vctk-1rnd} selects one random utterance from VCTK; \emph{vctk-4rnd} selects four random utterances from VCTK; \emph{cross-ds-4rnd} samples one utterance from each dataset (VCTK, VoxCeleb1, CREMA-D, ESD); \emph{cremad-emo-4rnd} samples four CREMA-D utterances with specific emotions (angry, neutral, sad, happy).  Utterances in multi-prompt strategies randomly cropped to 3 seconds each to maintain total duration under 12 seconds.

\section{Results and Discussion}

\vspace{-4pt}\subsection{Model comparison}
We benchmark our model against the SOTA streaming system \emph{DarkStream} and offline approach \emph{EASY}. We compare with two \emph{DarkStream}'s configurations (Mel+CL and Wave+CL) at LA=140 as these match our model's latency, since lower LA for \emph{DarkStream} causes even worse WER degradation. The results are shown in Table \ref{tab:stream-anon-merged}. Compared with \emph{DarkStream}, our model improves intelligibility, improving WER relatively by up to 46\%  at comparable latency budgets. For emotion recognition, our emotion-prompt conditioning boosts expressiveness by $\sim$28\% and the other prompt strategy yield gains of 15\%. Privacy results are comparable: our model is slightly better under the lazy-informed attacker, while \emph{DarkStream} is better under the semi-informed attacker. Finally, offline baselines highlight the remaining gap between online and offline anonymization, indicating room for future advances in streaming approaches.

\begin{table}[t]
\centering
\setlength{\tabcolsep}{3pt}%
\caption{Performance comparison of online and offline SA methods. Numbers in parentheses show relative improvement/degradation over the selected \emph{DarkStream model} ({\color{ForestGreen} green} indicates improvement, {\color{RubineRed} red} indicates degradation). For privacy fairness, we select models for comparison only if they achieve close to 50\% EER under the lazy-informed attacker scenario, indicating near-random guess performance by the attacker.}
\label{tab:stream-anon-merged}
\resizebox{\columnwidth}{!}{%
\begin{tabular}{l c c c c c}
\toprule[0.4ex]
\textbf{Model} & \textbf{Type} &
\makecell[c]{\textbf{WER} $\downarrow$} &
\makecell[c]{\textbf{UAR} $\uparrow$} &
\makecell[c]{\textbf{EER} $\uparrow$\\\textbf{lazy-informed}} &
\makecell[c]{\textbf{EER} $\uparrow$\\\textbf{semi-informed}} \\
\midrule[0.2ex]
\emph{EASY} \cite{yao25_interspeech}    & Offline & 2.70 & 63.81 & -- & 45.89      \\
\midrule[0.2ex]
\multicolumn{6}{l}{\emph{DarkStream} \cite{quamer2025darkstream}}\\
\makecell[l]{Mel+CL,\,LA=140, KM} & \makecell[l]{Online,\\210\,ms} 
 & 8.75\,$_{ \color{gray} \nobreak (0.0\%)\,}$ & 34.73\,$_{ \color{gray} (0.0\%)\,}$ & 47.26\,$_{ \color{gray} (0.0\%)\,}$ & 21.83\,$_{ \color{gray} (0.0\%)\,}$   \\
\makecell[l]{Wave+CL,\,LA=140, KM} & \makecell[l]
{Online,\\200\,ms}
  & 9.52$_{\color{RubineRed} (8.8\% \text{\textuparrow})}$ & 34.49$_{\tiny \color{RubineRed} (0.7\% \text{\textdownarrow})}$ & 46.75$_{\tiny \color{RubineRed} (1.1\% \text{\textdownarrow})}$ & \textbf{22.68$_{\tiny \color{ForestGreen} (3.9\% \text{\textuparrow})}$}   \\  
\midrule[0.2ex]
\multicolumn{6}{l}{\emph{Stream-Voice-Anon} (Ours)} \\
\makecell[l]{cremad-emo-4rnd} &
\makecell[l]{Online,\\180\,ms} & 6.59$_{\tiny \color{ForestGreen} (24.7\% \text{\textdownarrow})}$ & \textbf{44.59$_{\tiny \color{ForestGreen} (28.4\% \text{\textuparrow})}$} & 46.53$_{ \color{RubineRed} (1.5\% \text{\textdownarrow})}$ & 18.63$_{ \color{RubineRed} (14.6\% \text{\textdownarrow})}$   \\
\makecell[l]{cross-ds-4rnd} &
\makecell[l]{Online,\\180\,ms} & \textbf{4.71$_{\tiny \color{ForestGreen} (46.2\% \text{\textdownarrow})}$} & 39.94$_{\tiny \color{ForestGreen} (15.0\% \text{\textuparrow})}$ & \textbf{47.72$_{ \color{ForestGreen} (0.9\% \text{\textuparrow})}$} & 18.98$_{ \color{RubineRed} (13.1\% \text{\textdownarrow})}$   \\
\bottomrule[0.4ex]
\end{tabular}}
\vspace*{-\baselineskip}

\end{table}

\vspace{-4pt}\subsection{Prompt diversity}

In this experiment, we study how prompt (target) conditioning diversity affects privacy, emotion preservation and intelligibility. Our motivation for increasing prompt diversity is that a semi-informed attacker can adapt by finetuning to an anonymization strategy, so varying the prompts is crucial to mask source speaker identity cues. As shown in Table \ref{tab:selection_strategy_ablation} increasing prompt diversity (\textit{vctk-1fix}\textrightarrow \textit{vctk-4rnd}\textrightarrow\textit{cross-ds-4rnd}) consistently raises EER against semi-informed attackers, supporting the hypothesis that diversity hinders attacker adaptation to anonymization techniques. Gains under lazy-informed threat are relatively smaller as the attacker does not adapt to anonymization techniques. Even a single fixed prompt utterance already provides anonymization close to random guess under that threat model.

\begin{table}[h]
\footnotesize
\centering
\setlength{\tabcolsep}{3pt}%
\caption{Impact of prompt selection strategies on anonymization performance and utility preservation.}
\label{tab:selection_strategy_ablation}
\begin{tabular}{l c c c c}
\toprule
\makecell[l]{\textbf{Selection}\\\textbf{strategy}} &
\makecell[c]{\textbf{WER} $\downarrow$} &
\makecell[c]{\textbf{UAR} $\uparrow$} &
\makecell[c]{\textbf{EER} $\uparrow$\\\textbf{lazy-inform}} &
\makecell[c]{\textbf{EER} $\uparrow$\\\textbf{semi-inform}} \\
\midrule[0.15ex]
\makecell[l]{\emph{vctk-1fix}}  & \textbf{4.54} & 39.71 & 47.19 & 15.92   \\
\makecell[l]{\emph{vctk-1rnd}}  & 4.70 & \textbf{40.55} & 45.88 & 15.00   \\
\makecell[l]{\emph{vctk-4rnd}}  & 4.74 & 40.36 & 44.96 & 16.35  \\
\makecell[l]{\emph{cross-ds-4rnd}} & 4.71 & 39.94 & \textbf{47.72} & \textbf{18.98}   \\
\bottomrule
\end{tabular}
\vspace*{-\baselineskip}

\end{table}

\vspace{-4pt}\subsection{Latency vs privacy vs utility trade-off}

\begin{figure}[]
\centerline{\includegraphics[width=1.05\linewidth]{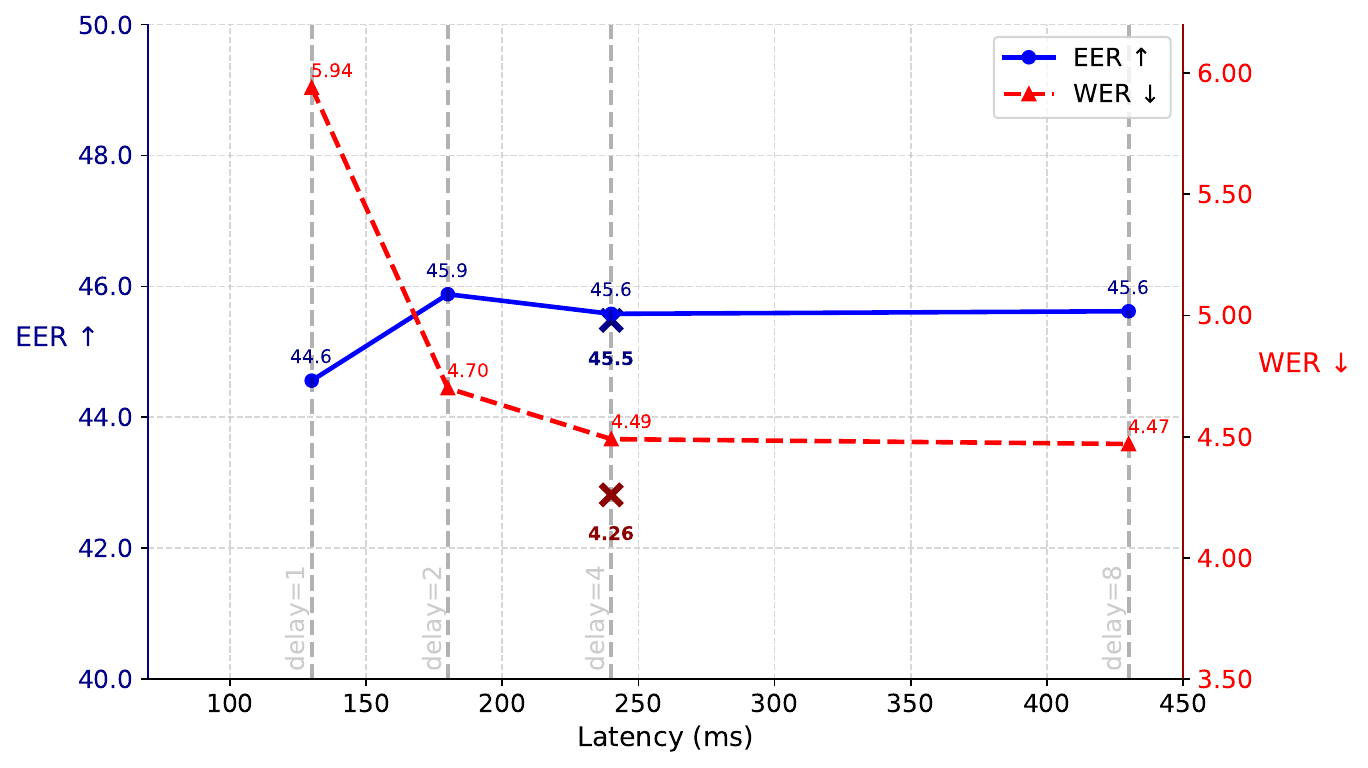}}
\caption{Latency (x-axis) against privacy (\textcolor{blue}{EER $\uparrow$}, left y-axis) and intelligibility (\textcolor{red}{WER $\downarrow$}, right y-axis) for \emph{Stream-Voice-Anon} with $d\in\{1, 2, 4, 8\}$ dynamic delay values. Bold \textcolor{BlueViolet}{$\pmb{\times}$}, \textcolor{BrickRed}{$\pmb{\times}$} markers indicate a fixed-delay model with $d=4$ for comparison. Selection strategy is \emph{vctk-1rnd}.}
\label{fig:LatencyvsEERvsWER}
\end{figure}

This experiment shows latency vs privacy vs utility trade-off for different latency budgets on Figure \ref{fig:LatencyvsEERvsWER}. Across 130-440ms, privacy is essentially invariant to latency, showing that it does not weaken protection. As expected, intelligibility improves with higher latency up to 180 ms and then flattens as the delay increases further. We then compare the dynamic delay model with a fixed delay at 240 ms latency ($d=4$), which achieves slightly lower WER but shows no privacy boost. Therefore, while fixed delay offers a small ASR advantage, dynamic delay preserves privacy and allows to choose latency at inference time without retraining to meet different requirements.

\vspace{-4pt}\subsection{Latency and RTF}

\begin{table}[t]
\centering
\normalsize

\setlength{\tabcolsep}{3pt}
\caption{Latency and real-time factor (RTF) performance across different chunk sizes on server and laptop hardware configurations with $d=2$.}
\resizebox{\columnwidth}{!}{
\begin{tabular}{cccc ccc}
\toprule[0.3ex]
& \multicolumn{3}{c}{Server (H200)} & \multicolumn{3}{c}{PC (Laptop 3060)} \\
\cmidrule(lr){2-4}\cmidrule(lr){5-7}
\textbf{Chunk size} & \textbf{Inference time} & \textbf{RTF} & \textbf{Latency} &
\textbf{Inference time} & \textbf{RTF} & \textbf{Latency} \\
\midrule
46\,ms  & 13\,ms & 0.28 & 151\,ms & 43\,ms & 0.93 & 180\,ms \\
92\,ms  & 17\,ms & 0.18 & 201\,ms & 53\,ms & 0.58 & 237\,ms \\
276\,ms & 31\,ms & 0.11 & 399\,ms & 96\,ms & 0.35 & 464\,ms \\
\bottomrule[0.3ex]
\end{tabular}}
\label{tab:latency_rtf}
\vspace*{-\baselineskip}
\end{table}

Table \ref{tab:latency_rtf} reports latency and RTF on the H200 server GPU and laptop RTX 3060. In H200, all settings are run in real time (RTF = 0.28 $\rightarrow$ 0.11) with 151-399 ms latency. On laptop GPU, RTF improves from 0.93 to 0.35 with 180-464 ms latency. Increasing chunk size reduces RTF but increases latency, revealing a throughput–responsiveness trade-off. For applications prioritizing responsiveness, 46–92 ms chunks yield the lowest delays, whereas 276 ms chunk offer the largest real-time margin when slightly higher latency is acceptable.

\section{Conclusion}
We present \emph{Stream-Voice-Anon}, a real-time SA system that successfully adapts LM-based VC architectures for streaming privacy protection. Our system demonstrates that NAC-based language modeling can be effectively extended to SA tasks while maintaining real-time performance. Compared to the SOTA streaming baseline \emph{DarkStream}, our system achieves substantial improvements in intelligibility (46\% WER relative reduction) and emotion preservation (28\% relative improvement) while maintaining comparable privacy protection against lazy-informed attackers (47.26\% vs 47.72\%) and latency (200\,ms vs 180\,ms). Against semi-informed attackers, our system shows slight degradation in privacy (21.83\% vs 18.98\% EER), indicating room for improvement against adaptive threats. The dynamic delay mechanism enables flexible latency-quality trade-offs without retraining, making it practical for various applications. However, the system currently requires GPU acceleration and cannot operate in real-time on CPU-only hardware. Future work will focus on improving disentanglement techniques to narrow the gap with offline methods, enhancing robustness against semi-informed attackers, optimizing for CPU deployment to enable broader accessibility.

\section{Acknowledgement}

This research was partially conducted using computing resources from the College of Computing and Data Science (CCDS), Nanyang Technological University (NTU).

{\footnotesize
\bibliographystyle{IEEEbib}
\bibliography{strings,refs}
}
\end{document}